\documentstyle[aps,eqsecnum,preprint,floats,epsf,epsfig]{revtex}
\begin{document}
\def\be{\begin{eqnarray}}
\def\en{\end{eqnarray}}
\def\non{\nonumber}
\def\la{\langle}
\def\ra{\rangle}
\def\nc{N_c^{\rm eff}}
\def\vp{\varepsilon}
\def\drho{\bar\rho}
\def\deta{\bar\eta}
\def\vma{{_{V-A}}}
\def\vpa{{_{V+A}}}
\def\J{{J/\psi}}
\def\K{{K^*}}
\def\dxi{\bar\xi}
\def\ov{\overline}
\def\Re{{\rm Re}}
\def\Im{{\rm Im}}
\def\Lqcd{{\Lambda_{\rm QCD}}}
\def\pr{{\sl Phys. Rev.}~}
\def\prl{{\sl Phys. Rev. Lett.}~}
\def\pl{{\sl Phys. Lett.}~}
\def\np{{\sl Nucl. Phys.}~}
\def\zp{{\sl Z. Phys.}~}
\def\lsim{ {\ \lower-1.2pt\vbox{\hbox{\rlap{$<$}\lower5pt\vbox{\hbox{$\sim$}
}}}\ } }
\def\gsim{ {\ \lower-1.2pt\vbox{\hbox{\rlap{$>$}\lower5pt\vbox{\hbox{$\sim$}
}}}\ } }


\begin {flushright}
\hspace{7in}
\parbox{1.5in}{
IPAS-HEP-01-007\\
KIAS-P01050\\}
\end {flushright}

\centerline{\Large\bf $B\to J/\psi K^*$ Decays in QCD
Factorization}
\bigskip \bigskip
\centerline{\bf Hai-Yang Cheng$^{1,2,3}$, Yong-Yeon Keum$^1$ and
Kwei-Chou Yang$^{4}$}
\medskip
\centerline{$^1$ Institute of Physics, Academia Sinica}
\centerline{Taipei, Taiwan 115, Republic of China}
\medskip
\centerline{$^2$ Physics Department, Brookhaven National
Laboratory} \centerline{Upton, New York 11973}
\medskip
\centerline{$^3$ C.N. Yang Institute for Theoretical Physics,
State University of New York} \centerline{Stony Brook, New York
11794}
\medskip
\centerline{$^4$ Department of Physics, Chung Yuan Christian
University} \centerline{Chung-Li, Taiwan 320, Republic of China}

\bigskip \bigskip

\bigskip
 \bigskip
\centerline{\bf Abstract}
\bigskip
{\small The hadronic decay $B\to\J K^*$ is analyzed within the
framework of QCD factorization. The spin amplitudes $A_0$, $A_\|$
and $A_\bot$ in the transversity basis and their relative phases
are studied using various different form-factor models for $B-K^*$
transition. The effective parameters $a_2^h$ for helicity
$h=0,+,-$ states receive different nonfactorizable contributions
and hence they are helicity dependent, contrary to naive
factorization where $a_2^h$ are universal and polarization
independent. QCD factorization breaks down even at the twist-2
level for transverse hard spectator interactions. Although a
nontrivial strong phase for the $A_\|$ amplitude can be achieved
by adjusting the phase of an infrared divergent contribution, the
present QCD factorization calculation cannot say anything definite
about the phase $\phi_\|$. Unlike $B\to\J K$ decays, the
longitudinal parameter $a_2^0$ for $B\to \J K^*$ does not receive
twist-3 corrections and is not large enough to account for the
observed branching ratio and the fraction of longitudinal
polarization. Possible enhancement mechanisms for $a_2^0$ are
discussed.

} \pagebreak

\section{Introduction}
It has been well known that the factorization approach (naive or
generalized) fails to explain the production ratio $R={\cal
B}(B\to\J K^*)/{\cal B}(B\to\J K)$ and the fraction of
longitudinal polarization $\Gamma_L/\Gamma$ in $B\to\J K^*$ decay.
We consider two representative form-factor models for $B-K(K^*)$
transitions, the Ball-Braun (BB) model based on the light-cone sum
rule (LCSR) analysis \cite{Ball} and the Melikhov-Stech (MS) model
\cite{Melikhov} based on the constituent quark picture. Both are
consistent with the lattice calculation at large $q^2$, the
constraint from $B\to\phi K^*$ at lower $q^2$ and the constraint
from heavy quark symmetry on the $q^2$ dependence of heavy-light
transition (see Sec. IV for more details). We see from Table I
that in general the predicted longitudinal polarization is too
small, whereas the production ratio is too large.

\begin{table}[ht]
\caption{The ratio of vector meson to pseudoscalar production $R$
and the longitudinal polarization fraction $\Gamma_L/\Gamma$ in
$B\to J/\psi K^{(*)}$ decays calculated in two representative
form-factor models using the factorization hypothesis.
 \label{tab:rgp}}
\begin{center}
\footnotesize
\begin{tabular}{lcccccc}
& & & \multicolumn{4}{c}{Experiments}  \\ \cline{4-7}
 & \raisebox{1.5ex}[0cm][0cm]{BB} & \raisebox{1.5ex}[0cm][0cm]{MS}
& CDF \cite{CDF} & CLEO \cite{CLEO} & BaBar \cite{BaBar} & Belle \cite{Belle} \\
\hline
  $R$  & 3.40 & 3.11 & $1.53\pm 0.32$ & $1.45\pm 0.26$ & $1.38\pm 0.11$ & $1.43\pm0.13$ \\
$\Gamma_L/\Gamma$& 0.47 & 0.46 & $0.61\pm 0.14$ & $0.52\pm 0.08$
 & $0.60\pm 0.04$ & $0.60\pm0.05$ \\
\end{tabular}
\end{center}
\end{table}

This is understandable because the parameter $a_2$, which governs
$B\to \J K (K^*)$ decays, is assumed to be universal according to
the factorization hypothesis, namely $a_2^{h}(\J K^*)=a_2(\J K)$
where $h=0,+,-$  refer to the helicity states $00$, $++$ and $--$
respectively. In the above-mentioned form-factor models, one has
$h_0=5.98$, $h_+=6.23$ and $h_-=043$ (in units of GeV$^3$) in the
BB model and $h_0=5.47$, $h_+=5.92$ and $h_-=0.73$ in the MS
model, where $h_i$ are the helicity amplitudes given by
 \be
h_{0} &=& {f_\J\over
2m_\K}\left[(m_B^2-m_{\J}^2-m_{\K}^2)(m_B+m_{\K})A^{B\K}_1(m_\J^2)
-{4m_B^2p_c^2\over m_B+m_\K }\,A^{B\K}_2(m_\J^2)\right], \non \\
h_{\pm} &=& m_\J f_\J\left[(m_B+m_{\K})A^{B\K}_1(m_\J^2)\pm
{2m_Bp_c\over m_B+m_\K }\,V^{B\K}(m_\J^2)\right].
 \en
It is obvious that $h_+> h_0\gg h_-$. Therefore, under naive
factorization $\Gamma_L/\Gamma\approx
(a_2^0h_0)^2/[(a_2^0h_0)^2+(a_2^+h_+)^2]= h^2_0/(h^2_0+h^2_+)\lsim
1/2$ and $R$ is expected to be greater than unity due to three
polarization states for $\J K^*$. These two problems will be
circumvented if nonfactorized terms contribute differently to each
helicity amplitude and to different decay modes so that $a_2^0(\J
K^*)> a_2^+(\J K^*)\neq a_2^-(\J K^*)$ and $a_2(\J K)> a_2^h(\J
K^*)$. In other words, the present data imply that the effective
parameter $a_2^h$ should be non-universal and polarization
dependent. Recently two of us have analyzed charmless $B\to VV$
decays within the framework of QCD factorization \cite{CYvv}. We
show that, contrary to phenomenological generalized factorization,
nonfactorizable corrections to each partial-wave or helicity
amplitude are not the same; the effective parameters $a_i$ vary
for different helicity amplitudes. The purpose of the present
paper is to study the nonfactorizable effects in $B\to \J K^*$
decay within the same framework of QCD factorization.

The decays $B\to \J K(K^*)$ are of great interest as
experimentally only a few color suppressed modes in hadronic $B$
decays have been measured so far. The recent measurement by BaBar
\cite{BaBar} has confirmed the earlier  CDF observation \cite{CDF}
that there is a nontrivial strong phase difference between
polarized amplitudes, indicating final-state interactions.
However, no such evidence is seen by CLEO \cite{CLEO} and more
recently by Belle \cite{Belle}. It is interesting to check if the
current approach for $B$ hadronic decays predicts a departure from
factorization. Therefore, the measurements of various helicity
amplitudes in $B\to\J K^*$ decays will provide a nice ground for
testing factorization and differentiating various theory
approaches in which the calculated nonfactorizable terms have real
and imaginary parts.

It is known that in the QCD factorization approach the coefficient
$a_2$ is severely suppressed in the absence of hard spectator
interactions. It has been shown in \cite{CYJpsiK} that $|a_2|$ in
$B\to \J K$ is of order 0.11 to the leading twist order, to be
compared with the experimental value of order 0.25. The twist-3
effect in hard spectator interactions will enhance $a_2$ to the
value of $0.19^{+0.14}_{-0.12}$. We shall see later that, contrary
to the $\J K$ case, $a_2^0$ in $B\to\J K^*$ does not receive
twist-3 contributions and it is dominated by twist-2 hard
spectator interactions.

The layout of the present paper is as follows. In Sec. II we first
outline the necessary ingredients of the QCD factorization
approach for describing $B\to \J K^*$ and then we proceed to
compute vertex and hard spectator interactions. The ambiguity of
the experimental determination of spin amplitude phases is
addressed in Sec. III. Numerical calculations and results are
presented in Sec. IV. Discussions and conclusions are shown in
Sec. V.

\section{$B\to\J K^*$ in QCD Factorization}
\subsection{Factorization formula}
The general $B\to\J K^*$ amplitude consists of three independent
Lorentz scalars:
 \be A[B(p)\to
\J(\vp_\J,p_\J)\K(\vp_\K,p_\K)] \propto
\vp_\J^{*\mu}\vp_\K^{*\nu}(a g_{\mu\nu}+bp_\mu
p_\nu+ic\epsilon_{\mu\nu\alpha\beta}p_\J^\alpha p_\K^\beta),
\label{amp}
 \en
where $\epsilon^{0123}=+1$ in our convention, the coefficient $c$
corresponds to the $P$-wave amplitude, and $a$, $b$ to the mixture
of $S$- and $D$-wave amplitudes. Three helicity amplitudes can be
constructed as \footnote{For $\ov B\to\J \ov K^*$ decay the
transverse amplitudes are given by $H_\pm=-a\pm m_Bp_c\,c$.}
 \be
H_{0} &=& -{1\over 2m_\J m_\K}\left[
(m_B^2-m_\J^2-m_\K^2)a+2m_B^2p_c^2b\right], \non \\ H_{\pm} &=&
a\pm m_Bp_c\,c, \label{Hii}
 \en
where $p_c$ is the c.m. momentum of the vector meson in the $B$
rest frame.  If the final-state two vector mesons are both light
as in charmless $B\to V_1V_2$ decays with $V_1$ being a recoiled
meson and $V_2$ an ejected one, it is expected that
$|H_{0}|^2>|H_+|^2>|H_-|^2$ owing to the argument that the
amplitude $H_+$ is suppressed by a factor of $\sqrt{2}m_2/m_B$ as
one of the quark helicities in $V_2$ has to be flipped, while the
$H_-$ amplitude is subject to a further chirality suppression of
order $m_1/m_B$ \cite{Korner}. However, for $B\to\J K^*$ decay,
$\sqrt{2}m_\J/m_B$ is of order unity and hence in practice $H_+$
and $H_0$ can be comparable.

Note that the polarized decay amplitudes can be expressed in
several different but equivalent bases. For example, the helicity
amplitudes can be related to the spin amplitudes in the
transversity basis $(A_0, A_\|, A_\bot)$ defined in terms of the
linear polarization of the vector mesons, or to the partial-wave
amplitudes $(S,P,D)$ via:
 \be
A_0 &=& H_{0}= -{1\over\sqrt{3}}\, S+\sqrt{2\over 3}\, D, \non \\
A_\| &=& {1\over\sqrt{2}}(H_{+}+ H_{-})=\sqrt{2\over 3}
\,S+{1\over\sqrt{3}}\,D, \non \\
A_\bot &=& {1\over\sqrt{2}}(H_{+}- H_{-})=P,
 \en
where we have followed the sign convention of \cite{Dighe}. The
decay rate reads
 \be
\Gamma(B\to \J K^*) &=& {p_c\over
8\pi m_B^2}|{G_F\over\sqrt{2}}V_{cb}V_{cs}^*|^2(|H_{0}|^2+|H_{+}|^2+|H_{-}|^2), \non \\
&=& {p_c\over 8\pi m_B^2}|{G_F\over\sqrt{2}}V_{cb}V_{cs}^*|^2(|A_{0}|^2+|A_\bot|^2+|A_\||^2), \non \\
&=& {p_c\over 8\pi
m_B^2}|{G_F\over\sqrt{2}}V_{cb}V_{cs}^*|^2(|S|^2+|P|^2+|D|^2). \en

The effective Hamiltonian relevant for $B\to \J K^*$  has the form
\be \label{hamiltonian} {\cal H}_{\rm eff} =
{G_F\over\sqrt{2}}\Bigg\{V_{cb}V_{cs}^*
\Big[c_1(\mu)O_1(\mu)+c_2(\mu)O_2(\mu)\Big]
-V_{tb}V_{ts}^*\sum^{10}_{i=3}c_i(\mu)O_i(\mu)\Bigg\}+{\rm h.c.},
\en where \be
 && O_1= (\bar cb)_\vma(\bar sc)_\vma,
\qquad\qquad\qquad\qquad\quad~ O_2 = (\bar sb)_\vma(\bar cc)_\vma,
\non
\\ && O_{3(5)}=(\bar sb)_\vma\sum_{q'}(\bar q'q')_{\vma(\vpa)},
\qquad  \qquad~ O_{4(6)}=(\bar s_\alpha
b_\beta)_\vma\sum_{q'}(\bar q'_\beta q'_\alpha)_{ \vma(\vpa)},
\\ && O_{7(9)}={3\over 2}(\bar sb)_\vma\sum_{q'}e_{q'}(\bar
q'q')_{\vpa(\vma)},
  \qquad~ O_{8(10)}={3\over 2}(\bar s_\alpha b_\beta)_\vma\sum_{q'}e_{q'}(\bar
q'_\beta q'_\alpha)_{\vpa(\vma)},   \non
 \en
with $O_3$--$O_6$ being the QCD penguin operators,
$O_{7}$--$O_{10}$ the electroweak penguin operators, and $(\bar
q_1 q_2)_{_{V\pm A}}\equiv\bar q_1\gamma_\mu(1\pm\gamma_5)q_2$.
Under factorization, the decay amplitude of $B\to\J K^*$ reads
 \be
A(B\to\J
K^*)=\,{G_F\over\sqrt{2}}V_{cb}V_{cs}^*(a_2+a_3+a_5+a_7+a_9)X^{(
BK^*,\J)},
 \en
where
 \be
 X^{( BK^*,\J)} &\equiv & \la
\J|(\bar cc)_\vma|0\ra\la K^*|(\bar b s)_\vma|B\ra \non \\ &=& -
if_\J m_\J\Bigg[(\vp^*_{K^*}\cdot\vp^*_\J) (m_{B}+m_{K^*})A_1^{ BK^*}(m_{\J}^2)  \non \\
&-& (\vp^*_{K^*}\cdot p_{_{B}})(\vp^*_\J \cdot p_{_{B}}){2A_2^{
BK^*}(m_{\J}^2)\over m_{B}+m_{K^*} } -
i\epsilon_{\mu\nu\alpha\beta}\vp^{*\mu}_\J\vp^{*\nu}_{K^*}p^\alpha_{_{B}}
p^\beta_{K^*}\,{2V^{ BK^*}(m_{\J}^2)\over m_{B}+m_{K^*} }\Bigg].
\label{factamp}
 \en
Note that for $\ov B\to\J\ov K^*$ decay, the factorizable
amplitude  $X^{( \ov B\ov K^*,\J)} \equiv \la \J|(\bar
cc)_\vma|0\ra\la \ov K^*|(\bar s b)_\vma|\ov B\ra$ is the same as
(\ref{factamp}) except that the last term proportional to
$i\epsilon_{\mu\nu\alpha\beta}$ has a positive sign. Comparing
(\ref{factamp}) with  (\ref{Hii}) leads to the helicity amplitudes
 \be
 H_{0} = -\tilde a(\J K^*)h_0,\qquad H_{\pm} = \tilde a(\J K^*)
 h_\pm,
 \en
where $\tilde a(\J K^*)=a_2+a_3+a_5+a_7+a_9$. Note that the
helicity amplitudes $H_\pm$ in $\ov B\to\J \ov K^*$ are precisely
the ones $H_\mp$ in $B\to\J K^*$ decays. Hence, in the
factorization approach one has $|H_-|>|H_+|$ for the former and
$|H_+|>|H_-|$ for the latter. This is consistent with the picture
that the $s$ quark produced in the weak process $b\to c\bar c s$
in $\ov B\to\J \ov K^*$ has helicity $-1/2$ in the zero quark mass
limit. Therefore, the helicity of $\ov K^*$ in $\ov B\to\J \ov
K^*$ cannot be +1 and the corresponding helicity amplitude $H_+$
vanishes in the chiral limit \cite{Suzuki}.

\subsection{QCD factorization}
Under naive factorization, the coefficients $a_i$ are given by
$a_{2i}=c_{2i}+{1\over N_c}c_{2i-1},~ a_{2i-1}=c_{2i-1}+{1\over
N_c}c_{2i}$. Hence, $a_2^{h}(\J K^*)=a_2(\J K)$ for $h=0,+,-$. In
the present paper, we will compute nonfactorizable corrections to
$a_2^h(\J K^*)$. The effective parameters $a_i^h$ entering into
the helicity amplitudes $H_{0}$ and $H_{\pm}$ are not the same.

The QCD-improved factorization approach advocated recently in
\cite{BBNS} allows us to compute the nonfactorizable corrections
in the heavy quark limit since only hard interactions between the
$(BV_1)$ system and $V_2$ survive in the $m_b\to\infty$ limit.
Naive factorization is recovered in the heavy quark limit and to
the zeroth order of QCD corrections. In this approach, the
light-cone distribution amplitudes (LCDAs) play an essential role.
The LCDAs of the vector meson are given by \cite{Ballv,BBNS}
 \be \la V(P,\vp)|\bar q(x)\gamma_\mu
q'(0)|0\ra  &=& f_V m_V\int^1_0 d\xi\,e^{i\xi P\cdot
x}\Bigg[{\vp^{*}\cdot x\over P\cdot x} P_\mu\Phi^V_\|(\xi)
+(\vp^{*}_\mu-{\vp^{*}\cdot
x\over P\cdot x} P_\mu)g_\bot^{(v)}(\xi)\Bigg], \non \\
\la V(P,\vp)|\bar q(x)\gamma_\mu\gamma_5 q'(0)|0\ra &=& {1\over
4}m_V\left(f_V-f_V^T\,{m_q+m_{q'}\over m_V}\right)
\epsilon_{\mu\nu\alpha\beta}\vp^{*\nu}P^\alpha
x^\beta\int^1_0 d\xi\,e^{i\xi P\cdot x}g_\bot^{(a)}(\xi), \non \\
\la V(P,\vp)|\bar q(x)\sigma_{\mu\nu}q'(0)|0\ra &=& -if_V^T
(\vp^{*}_\mu P_\nu-\vp^{*}_\nu P_\mu)\int^1_0
d\xi\,e^{i\xi P\cdot x}\Phi^V_\bot(\xi) \non \\
 && -if^T_V (P_\mu x_\nu-P_\nu x_\mu){\vp^{*}\cdot x\over (P\cdot
x)^2}m_V^2\int^1_0 d\xi\,e^{i\xi P\cdot x}h_\|^{(t)}(\xi), \non \\
\la V(P,\vp)|\bar q(x) q'(0)|0\ra &=& i{1\over 2}
\left(f_V-f_V^T\,{m_q+m_{q'}\over m_V}\right)(\vp^{*}\cdot
x)m_V^2\int^1_0 d\xi\,e^{i\xi P\cdot x}h_\|^{(s)}(\xi),
 \label{Vwf}
 \en
where $x^2=0$, $\xi$ is the light-cone momentum fraction of the
quark $q$ in the vector meson, $f_V$ and $f^T_V$ are vector and
tensor decay constants, respectively, but the latter is scale
dependent. In Eq. (\ref{Vwf}), $\Phi_\|(\xi)$ and $\Phi_\bot(\xi)$
are twist-2 DAs, while $h_\|^{(s,t)}$, $g_\bot^{(v)}$ and
$g_\bot^{(a)}$ are twist-3 ones. Since
 \be \label{polrel}
{\vp\cdot x\over P\cdot x}\,P^\mu=\vp^{\mu}_\|+{\vp\cdot x\over
P\cdot x}\,{m_V^2\over 2P\cdot x}\,x^\mu,
 \en
it is clear that to order ${\cal O}(m_V^2/m_B^2)$  the
approximated relation ${\vp\cdot x\over P\cdot
x}\,P^\mu=\vp^\mu_\|$ holds for a light vector meson, where
$\vp^\mu_\|$ ($\vp^\mu_\bot$) is the polarization vector of a
longitudinally (transversely) polarized vector meson. Also, to a
good approximation one has $\vp^\mu_\|=P^\mu_V/m_V$ for a light
vector meson like $K^*$, Hence, $P\cdot \vp_\bot=0$ and Eq.
(\ref{Vwf}) can be simplified for $K^*$ as
 \be
\la K^*(P,\vp)|\bar q(x)\gamma_\mu s(0)|0\ra &=& f_\K m_\K\int^1_0
d\xi\,e^{i\xi P\cdot x}\left[\vp^*_{\mu\|}
\Phi^\K_\|(\xi)+\vp^*_{\mu\bot} g_\bot^{K^*(v)}(\xi)\right], \non \\
\la\K(P,\vp)|\bar q(x)\gamma_\mu\gamma_5 s(0)|0\ra &=& {1\over
4}m_\K f_\K \epsilon_{\mu\nu\alpha\beta}\vp_\bot^{*\nu}P^\alpha
x^\beta\int^1_0 d\xi\,e^{i\xi P\cdot x}g_\bot^{K^*(a)}(\xi), \non \\
\la\K(P,\vp)|\bar q(x)\sigma_{\mu\nu}s(0)|0\ra &=& -if_\K^T
(\vp^*_{\mu\bot} P_\nu-\vp^{*}_{\nu\bot}
P_\mu)\int^1_0 d\xi\,e^{i\xi P\cdot x}\Phi^\K_\bot(\xi), \non \\
\la\K(P,\vp)|\bar q(x) s(0)|0\ra &=& -{1\over 2}f_\K m_\K \int^1_0
d\xi\,e^{i\xi P\cdot x}h_\|^{'(s)}(\xi), \label{K*wf}
 \en
where $h'(\xi)=dh(\xi)/d\xi$ and we have neglected light quark
masses and applied the relation
 \be
(P_\mu x_\nu-P_\nu x_\mu)\,{\vp\cdot x\over (P\cdot
x)^2}\,m_V^2={\vp\cdot x\over P\cdot x}(P_\mu P_\nu-P_\nu
P_\mu)+(\vp_{\mu\|} P_\nu-\vp_{\nu\|} P_\mu),
 \en
which vanishes for a light vector meson.  From Eq. (\ref{K*wf}) we
see that the twist-3 DA $h_\|^{(t)}$ of $K^*$ does not make a
contribution.

In the heavy quark limit, the $B$ meson wave function is given by
 \be
\la 0|\bar b_\alpha(x)q_\beta(0)|
B(p)\ra\!\!\mid_{x_+=x_\bot=0}=-{if_B\over 4}[(p\!\!\!/
+m_B)\gamma_5]_{\beta\gamma}\int^1_0d\drho\, e^{-i\drho
p_+x_-}[\Phi^B_1(\drho)+n\!\!\!/_-\Phi^B_2(\drho)]_{\gamma\alpha},
 \label{Bwf}
 \en
with  $n_-=(1,0,0,-1)$ and the normalization conditions
 \be
\int^1_0d\drho \,\Phi^B_1(\drho)=1, \qquad\quad \int^1_0d\drho\,
\Phi^B_2(\drho)=0.
 \en
Likewise, to the leading order in $1/m_c$, the $\J$ wave function
has a similar expression
 \be
\la \J(p,\vp)|\bar
c_\alpha(x)c_\beta(0)|0\ra\!\!\mid_{x_+=x_\bot=0} & &={f_\J\over
4}[\vp\!\!\!/^*(p\!\!\!/ +m_\J)]_{\beta\gamma} \non
\\ && \times \int^1_0d\xi\, e^{-i\xi
p_+x_-}[\Phi^\J_1(\xi)+n\!\!\!/_-\Phi^\J_2(\xi)]_{\gamma\alpha}.
 \label{Jwf}
 \en
Since the $\J$ meson is heavy, the use of the light-cone wave
function for $\J$ is problematic. The effects of higher twist wave
functions have to be included and may not converge fast enough.
Because the charmed quark in $J/\psi $ carries a momentum fraction
of order $\sim m_c/m_\J$, the distribution amplitudes of $J/\psi $
vanish in the end point region. In the following study we adopt
$\Phi_{||}$ as the DA of the non-local vector current of $\J$
rather than $g_\perp^{(v)}$ as the DA of the $\epsilon_\perp$
component since the latter does not vanish at the end point.
Hence, we will treat the $\J$ wave function on the same footing as
the $B$ meson. Comparing Eq.~(\ref{Jwf}) with Eq.~(\ref{Vwf}) we
see that at the leading order in $1/m_c$ one has
 \be
\Phi^\J_1(\xi)=\Phi^\J_\|(\xi)=\Phi^\J_\bot(\xi),\qquad
f_\J^T=f_\J. \label{wf&dc}
 \en

The inclusion of vertex-type corrections and hard spectator
interaction in QCD factorization leads to
 \be \label{ai}
 a_2^h &=& c_2+{c_1\over N_c}+{\alpha_s\over 4\pi}\,{C_F\over
N_c}c_1\,F^h, \non \\
 a_3^h &=& c_3+{c_4\over N_c}+{\alpha_s\over
4\pi}\,{C_F\over N_c} c_4\,F^h, \non \\
 a_5^h &=& c_5+{c_6\over
N_c}+{\alpha_s\over 4\pi}\,{C_F\over N_c} c_6(-F^h-12), \non \\
 a_7^h &=& c_7+{c_8\over N_c}+{\alpha_s\over 4\pi}\,{C_F\over N_c}
c_8(-F^h-12),  \non \\
 a_9^h &=& c_9+{c_{10}\over N_c}+{\alpha_s\over 4\pi}\,{C_F\over N_c}
 c_{10}\,F^h,
 \en
where $C_F=(N_c^2-1)/(2N_c)$ and the superscript $h$ denotes the
polarization of the vector mesons: $h=0$ for helicity 0 state, and
$h=\pm$ for helicity $\pm$ ones. In the naive dimensional
regularization (NDR) scheme for $\gamma_5$, $F^h$ in Eq.
(\ref{ai}) has the form
 \be
F^h=-12\ln{\mu\over m_b}-18+f_I^h+f_{II}^h, \label{F}
 \en
where the hard scattering function $f_I^h$ arises from vertex
corrections [see Figs. 1(a)-1(d)] and $f_{II}^h$ from the hard
spectator interactions with a hard gluon exchange between the
emitted vector meson and the spectator quark of the $B$ meson, as
depicted in Figs. 1(e)-1(f).

\begin{figure}[tb]
\vspace{-3.5cm}\psfig{figure=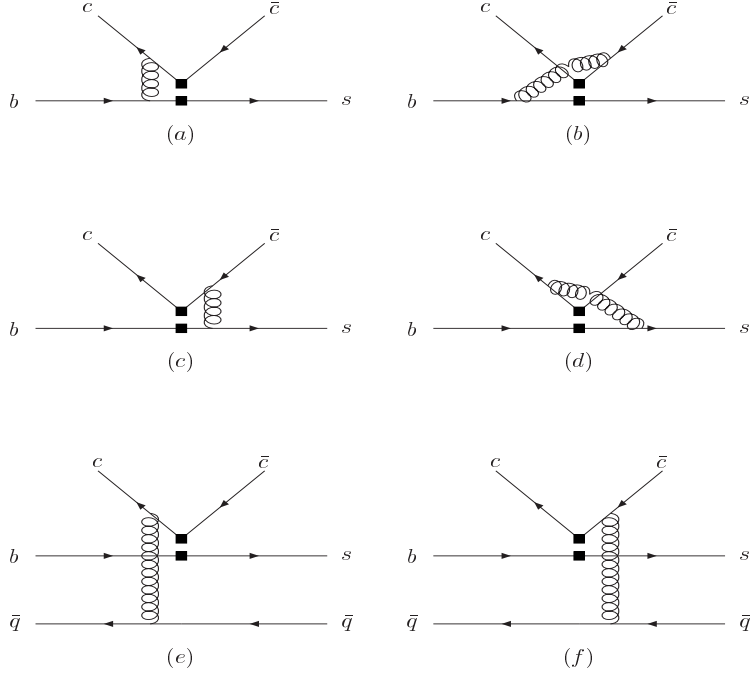,height=7.5 in,width=6.5in}
\vspace{-5.5cm}
    \caption{{\small Vertex and spectator corrections to $B\to J/\psi K^*$.}}
   \label{fig:vert}
\end{figure}

\subsection{Vertex corrections}
The calculation of vertex corrections in Fig. 1 is very similar to
that in $B\to\J K$ decay and the detail can be found in
\cite{CYJpsiK}. In terms of the two hard kernels $f_I$ and $g_I$
given by
 \be
f_I &=& \int^1_0 d\xi\,\Phi^\J_\|(\xi)\Bigg\{ {2z\xi\over
1-z(1-\xi)}+\left(3-2\xi\right){\ln \xi\over 1-\xi} \non
\\ &&+\left(-{3\over 1-z\xi}+{1\over
1-z(1-\xi)}-{2z\xi\over [(1-z(1-\xi)]^2}\right)z\xi\ln z\xi \non
\\ &&+\left(3(1-z)+2z\xi+{2z^2\xi^2\over
1-z(1-\xi)}\right){\ln (1-z)-i\pi\over 1-z(1-\xi)}\Bigg\} \non \\
&& +\int^1_0 d\xi\,\Phi^\J_\bot(\xi)\Bigg\{-4r{\ln \xi\over
1-\xi}+{4zr\ln z\xi\over 1-z(1-\xi)}-4zr{\ln (1-z)-i\pi\over
1-z(1-\xi)}\Bigg\}, \label{fI}
 \en
and
 \be
g_I &=& \int^1_0 d\xi\,\Phi^\J_\|(\xi)\Bigg\{{-4\xi\over
(1-z)(1-\xi)}\ln \xi+{z\xi\over [1-z(1-\xi)]^2}\ln (1-z)+
\Bigg({1\over (1-z\xi)^2}  \non \\ && -{1\over [1-z(1-\xi)]^2}
+{2(1+z-2z\xi)\over (1-z)(1-z\xi)^2}\Bigg)z\xi\ln
z\xi-i\pi\,{z\xi\over [1-z(1-\xi)]^2}\Bigg\} \non \\
&& +\int^1_0 d\xi\,\Phi^\J_\bot(\xi)\Bigg\{{4r\over
(1-z)(1-\xi)}\ln \xi-{4rz\over (1-z)(1-z\xi)}\ln z\xi\Bigg\},
 \label{gI}
 \en
where $r=f_\J^T m_c/(f_\J m_\J)$, $z\equiv m_\J^2/m_B^2$, the
first scattering function $f_I^h$ induced from vertex corrections
has the form
 \be
f_I^0 &=& f_I+g_I(1-z)\,{A^{BK^*}_0(m_\J^2)\over \tilde A^{BK^*}_3(m_\J^2)}, \non \\
f_I^\pm &=& f_I, \label{fh}
 \en
where
 \be
\tilde A_3(q^2)=\,{m_B+m_\K\over
2m_\K}A_1(q^2)-{m_B^2-m_\J^2+m_\K^2\over 2m_\K(m_B+m_\K)}A_2(q^2).
 \en
In writing Eqs. (\ref{fI}) and (\ref{gI}) we have distinguished
the contributions from $\Phi^\J_\|$ and $\Phi^\J_\bot$ for
reader's convenience, though later we will apply Eq.
(\ref{wf&dc}). Also notice that we have applied the
relation~\cite{CYJpsiK}\footnote{It is known from HQET that below
the $\overline m_c$ scale, where $\overline m_c$ is the pole mass
of the charmed quark, the vector and tensor currents receive the
same anomalous dimensions; that is, $f^T_{J/\psi}$ and
$f_{J/\psi}m_c$ scale as the same power. Up to the $m_b$ scale,
$f^T$ rescales with a factor $[\alpha_s(m_b)/\alpha_s(\overline
m_c)]^{4/(3b)}$, $m_c$ with $[\alpha_s(m_b)/\alpha_s(\overline
m_c)]^{4/b}$, and the ratio of $f^T_{J/\psi}/f_{J/\psi}$ becomes
$[\alpha_s(\overline m_c)/\alpha_s(m_b)]^{8/3b}$
$2m_c(m_b)/m_{J/\psi}=(1.1-1.2)\times 2m_c(m_b)/m_{J/\psi}$, where
$b=(11N_c-2n_f)/3$ and $m_c(m_b)$ is the running charmed quark
mass at the $m_b$ scale. However, the scale factor
$[\alpha_s(\overline m_c)/\alpha_s(m_b)]^{8/(3b)}\sim(1.1-1.2)$ is
relatively small and can be neglected for our purposes.}
 \be
r\equiv {f_\J^T m_c\over f_\J m_\J}=2\left({m_c\over
m_\J}\right)^2=2\xi^2. \label{relation}
 \en

Three remarks are in order. (i) As shown in \cite{CYJpsiK}, the
transverse DA $\Phi_\bot^\J$ contributes not only to the
transverse amplitudes $H_\pm$ but also to the longitudinal
amplitude $H_0$, and vice versa for the longitudinal DA
$\Phi_\|^\J$. This occurs because $\J$ is heavy: the coefficient
in front of $\Phi_\|$ in Eq. (\ref{polrel}) consists of not only
the longitudinal polarization but also the transverse one. (ii) It
is easily seen that in the zero $\J$ mass limit,
 \be f^0_I\to\int^1_0 d\xi\,\phi^\J(\xi)\left(3{1-2\xi\over
1-\xi}\ln\xi-3i\pi\right),
 \en
in agreement with  \cite{BBNS} for $B\to\pi\pi$, as it should be.
(iii) The expression of $A_0/\tilde A_3$ in Eq. (\ref{fh}) can be
further simplified by applying equations of motion. Neglecting the
mass of light quarks, applying the equation $\bar s
p\!\!\!/_\K(1-\gamma_5)b=0$ and sandwiching it between the $K^*$
and $B$ states leads to the result:
 \be
 -{m_\J^2\over 2m_B m_\K}A_2(q^2)=A_3(q^2)-A_0(q^2)
 \en
and hence $A_0^{BK^*}(m_\J^2)/ \tilde A_3^{BK^*}(m_\J^2)=1$.
Consequently, $f_I^0=f_I+g_I(1-z)$.

\subsection{Hard spectator interactions}
For hard spectator interactions, we write
 \be
f_{II}=f_{II(2)}+f_{II(3)},
 \en
where the subscript (...) denotes the twist dimension of the LCDA.
To the leading-twist order, we obtain
 \be f^{0}_{II(2)} &=& {4\pi^2\over
N_c}\,{\alpha_s(\mu_h)\over \alpha_s(\mu)}\,{f_Bf_\J f_\K \over
h_{0}}\,(1-z)\int^1_0 d\xi\,
d\drho\, d\deta\,\Phi^B_1(\drho)\Phi^\J(\xi)\Phi^\K(\deta)  \non \\
&\times & { \drho-\deta+(\drho-2\xi+\deta)z+4\xi^2z\over
\drho(\drho-\deta+\deta
z)[(\drho-\xi)(\drho-\deta)+(\deta\drho-\deta\xi-\drho\xi)z]}.
\label{fII2}
 \en
This can be further simplified by noting that $\drho\sim {\cal
O}(\Lqcd/m_b)\to 0$ in the $m_b\to\infty$ limit. Hence,
 \be
f^0_{II(2)} = {4\pi^2\over N_c}\,{\alpha_s(\mu_h)\over
\alpha_s(\mu)}\,{f_Bf_\J f_\K \over h_{0}}\int^1_0 d\drho\,
{\Phi^B_1(\drho)\over \drho}\int^1_0 d\xi \,{\Phi^\J(\xi)\over
\xi}\int^1_0 d\deta\, {\Phi^\K(\deta)\over \deta}, \label{fII2'}
 \en
where the $z$ terms in the numerator cancel after the integration
over $\xi$ via Eq. (\ref{relation}). Likewise, for transversely
polarization states, we find
 \be
f^\pm_{II(2)} &=& -{4\pi^2\over N_c}\,{\alpha_s(\mu_h)\over
\alpha_s(\mu)}\,{2f_Bf_\J f^\bot_\K m_\J\over m_Bh_{\pm}}(1\pm 1)
\non \\ &&\times \int^1_0 d\drho\,d\xi d\deta
\Phi^B_1(\drho)\Phi^\J(\xi)\Phi^\K_\bot(\deta)\,{1-2\xi\over
\drho\deta^2(1-z)}. \label{fII2pm}
 \en
Note that the hard gluon exchange in the spectator diagrams is not
as hard as in the vertex diagrams. Since the virtual gluon's
momentum squared there is $k^2=(-\bar \rho p_B+\bar\eta
p_\K)^2\approx -\bar\rho\bar\eta m_B^2\sim -\Lambda_h m_b$, where
$\Lambda_h$ is the hadronic scale $\sim 500$~MeV, we will set
$\alpha_s\approx \alpha_s(\sqrt{\Lambda_h m_b})$ in the spectator
diagrams. The corresponding Wilson coefficients in the spectator
diagrams are also evaluated at the $\mu_h=\sqrt{\Lambda_h m_b}$
scale. As for twist-3 contributions to hard spectator
interactions, we find
 \be
f_{II(3)}^0=0,
 \en
and
 \be \label{fII3}
f_{II(3)}^\pm &=& {4\pi^2\over N_c}\,{\alpha_s(\mu_h)\over
\alpha_s(\mu)}\,{2f_Bf_\J f_\K m_\J m_{K^*}\over m_B^2 h_{\pm}} \non \\
&& \times\int^1_0 d\drho\, {\Phi^B_1(\drho)\over \drho}\int^1_0
d\xi \,{\Phi^\J(\xi)\over \xi}\int^1_0
d\deta\,\left({g_\bot^{K^*(v)}(\deta)\over \deta(1-z)}\pm
{g_\bot^{K^*(a)}(\deta)\over 4\deta^2(1-z)}\right).
 \en

Since asymptotically $\Phi^\K(\deta)= 6\deta(1-\deta)$, the
logarithmic divergence of the $\deta$ integral in Eq.
(\ref{fII2'}) implies that the spectator interaction is dominated
by soft gluon exchanges between the spectator quark and the
charmed or anti-charmed quark of $\J$. Hence, QCD factorization
breaks down even at the twist-2 level for $f_{II(2)}^+$. Thus we
will treat the divergent integral as an unknown ``model" parameter
and write
 \be
 Y\equiv\int^1_0 {d\deta\over \deta}=\ln\left({m_B\over
\mu_h}\right)(1+\rho_H), \label{logdiv}
 \en
with $\rho_H$ being a complex number whose phase may be caused by
soft rescattering \cite{BBNS}. Note that linear divergences are
cancelled owing to the relation (\ref{relation}). Needless to say,
how to treat the unknown parameter $\rho_H$ is a major theoretical
uncertainty in the QCD factorization approach.

\subsection{Distribution amplitudes}
If we apply the asymptotic form for the vector meson's LCDAs
\cite{Ballv}
 \be && \Phi^V_\|(x)=\Phi^V_\bot(x)=g_\bot^{(a)}(x)=6x(1-x),  \non \\
&& g_\bot^{(v)}(x)={3\over 4}\left[ 1+(2x-1)^2\right],
 \en
it is easy to check that $f_{II(3)}^-=0$. Since the scale relevant
to hard spectator interactions is of order $\mu_h=\sqrt{\Lambda_h
m_b}\approx 1.5$ GeV, it is important to take into account the
evolution of LCDAs from $\mu=\infty$ down to the lower scale. The
leading-twist LCDA $\Phi_M$ can be expanded in terms of Gegenbauer
polynomials $C^{3/2}_n$ \cite{Ballv}:
 \be
\Phi_M(x,\mu)=6x(1-x)\left(1+\sum_{n=1}^\infty
a_{2n}^M(\mu)C_{2n}^{3/2}(2x-1)\right),
 \en
where the Gegenbauer moments $a_n^M$ are multiplicatively
renormalized. To $n=2$ we have
 \be
\Phi_\|^V(x,\mu) &=& 6x(1-x)\left[1+3a_1^\|\xi+{3\over
2}a_2^\|(5\xi^2-1)\right],  \non \\
\Phi_\bot^V(x,\mu) &=& 6x(1-x)\left[1+3a_1^\bot\xi+{3\over
2}a_2^\bot(5\xi^2-1)\right],
 \en
where $\xi=2x-1$. For twist-3 DAs we follow \cite{Ball98b} to
use\footnote{Note that there is a slight difference for the
expressions of $g_\bot^{(v,a)}$ in \cite{Ball98b} and
\cite{Ball98a}.}
 \be
 g_\bot^{(a)}(x,\mu) &=& 6x(1-x)\Bigg[1+a_1^\|\xi+\left\{{1\over
 4}a_2^\|+{5\over 3}\zeta_3\left(1-{3\over 16}\omega^A_3+{9\over
 16}\omega^V_3\right)\right\}(5\xi^2-1)\Bigg] \non\\
 &+& 6\delta_+[3x(1-x)+(1-x)\ln(1-x)+x\ln
 x] \non \\ &+& 6\delta_-[(1-x)\ln(1-x)-x\ln x],  \non \\
 g_\bot^{(v)}(x,\mu) &=& {3\over 4}(1+\xi^2)+{3\over
 2}a_1^\|\xi^3+\left({3\over 7}a_2^\|+5\zeta_3\right)(3\xi^2-1)
 \\
 &+& \left[{9\over 112}a_2^\|+{15\over
 64}\zeta_3\left(3\omega^V_3-\omega^A_3\right)\right](3-30\xi^2+35\xi^4)
 \non \\
 &+& {3\over 2}\delta_+[2+\ln x+\ln(1-x)]+{3\over
 2}\delta_-[2\xi+\ln(1-x)-\ln x],  \non
 \en
where the Gegenbauer moments and couplings $\eta_3$,
$\omega_3^{V,A}$, $\delta_{+,-}$ for $K^*$ at the scale $\mu^2=1$
GeV$^2$ and $\mu^2=5$ GeV$^2$ can be found in \cite{Ball98b}. It
turns out that the end-point behavior of $g_\bot^{(v)}$ for $K^*$
is substantially modified and is very different from that of the
asymptotic form (see Fig. 3 of \cite{Ball98a}).

\section{Experiments}
The angular analysis of $B^+\to\J K^{*+}$ and $B^0\to\J K^{*0}$
has been carried out by CDF \cite{CDF}, CLEO \cite{CLEO} and most
recently by the $B$ factories BaBar \cite{BaBar} and Belle
\cite{Belle}. The three polarized amplitudes are measured in the
transversity basis with results summarized in Table IV.
Experimental results are conventionally expressed in terms of spin
amplitudes $\hat A_{0,\bot,\|}$ normalized to unity, $|\hat
A_0|^2+|\hat A_\bot|^2+|\hat A_\||^2=1$. Since the measurement of
interference terms in the angular distribution is limited to
Re$(A_\|A_0^*)$, Im$(A_\perp A_0^*)$ and Im$(A_\perp A^*_\|)$,
there exists a phase ambiguity:
 \be
 \phi_\| & \to& -\phi_\|, \non \\
 \phi_\perp & \to& \pm\pi-\phi_\perp,  \\
 \phi_\perp-\phi_\| & \to& \pm \pi-(\phi_\perp-\phi_\|). \non
 \en
Take the BaBar measurement \cite{BaBar} as an example  :
 \be
 \phi_\perp=-0.17\pm0.17\,, \qquad\qquad \phi_\|=2.50\pm 0.22\,,
 \qquad \Rightarrow~~ |H_+|<|H_-|,
 \label{I}
 \en
where the phases are measured in radians. The other allowed
solution is
 \be
 \phi_\perp=-2.97\pm0.17\,, \qquad\qquad \phi_\|=-2.50\pm
 0.22\,,\qquad \Rightarrow~~ |H_+|>|H_-|.
 \label{II}
 \en
As pointed out in \cite{Suzuki}, the solution (\ref{I}) indicates
that $A_\|$ has a sign opposite to that of $A_\perp$ and hence
$|H_+|<|H_-|$, in contradiction to what expected from
factorization. Therefore, we will compare solution (\ref{II}) with
the factorization approach. Obviously there is a 3-$\sigma$ effect
that $\phi_\|$ is different from $\pi$ and this agrees with the
CDF measurement. However, such an effect is not observed by Belle
and CLEO (see Table IV). In Table IV we will only list those
amplitude phases from solution (\ref{II}).

The measured branching ratios are
 \be   \label{BRcharge}
 {\cal B}(B^+\to\J K^{*+})=\cases{ (13.7\pm 0.9\pm1.1)\times 10^{-4} &
 BaBar \cite{BaBar} \cr (12.9\pm0.8\pm1.2)\times 10^{-4} & Belle
 \cite{Belle} \cr (14.1\pm2.3\pm2.4)\times 10^{-4} & CLEO
 \cite{CLEO} \cr}
 \en
and
 \be \label{BRneutral}
 {\cal B}(B^0\to\J K^{*0})=\cases{ (12.4\pm0.5\pm0.9)\times 10^{-4} &
 BaBar \cite{BaBar} \cr (12.5\pm0.6\pm0.8)\times 10^{-4} & Belle
 \cite{Belle} \cr (13.2\pm1.7\pm1.7)\times 10^{-4} & CLEO
 \cite{CLEO} \cr}.
 \en

\section{Numerical results}
To proceed we use the next-to-leading Wilson coefficients in the
NDR scheme \cite{Buras96}
 \be
c_1=1.082, \quad c_2=-0.185,\quad c_3=0.014, \quad c_4=-0.035,
\quad c_5=0.009, \quad c_6=-0.041, \non \\
c_7/\alpha=-0.002, \quad c_8/\alpha=0.054, \quad
c_9/\alpha=-1.292, \quad c_{10}/\alpha=0.263,\quad c_g=-0.143,
 \en
at $\mu=\overline{ m}_b(m_b)=4.40$ GeV for
$\Lambda^{(5)}_{\overline{\rm MS}}=225$ MeV taken from Table XXII
of \cite{Buras96} with $\alpha$ being an electromagnetic
fine-structure coupling constant. For the decay constants, we use
 \be
f_{K^*}=221\,{\rm MeV}, \qquad f_\J=405\,{\rm MeV},\qquad
f_B=190\,{\rm MeV},
 \en
and we will assume $f^T_V=f_V$ for the tensor decay constant. For
LCDAs we use those in Sec.~II.E, and the $B$ meson wave function
 \be
\Phi^B_1(\drho)=N_B\drho^2(1-\drho)^2{\rm exp}\left[-{1\over
2}\left({\drho m_B\over \omega_B}\right)^2\right], \label{Bda}
 \en
with $\omega_B=0.25$ GeV and $N_B$ being a normalization constant.

In the following study, we will consider eight distinct
form-factor models: the Bauer-Stech-Wirbel (BSWI) model
\cite{BSW85,bsw}, the modified BSW model (referred to as the BSWII
model) \cite{NRSX}, the relativistic light-front (LF) quark model
\cite{cch}, the Neubert-Stech (NS) model \cite{ns}, the QCD sum
rule calculation by Yang \cite{yang}, the Ball-Braun (BB) model
based on the light-cone sum rule analysis \cite{Ball}, the
Melikhov-Stech (MS) model based on the constituent quark picture
\cite{Melikhov} and the Isgur-Wise scaling laws based on the SU(2)
heavy quark symmetry (YYK) so that the form factor $A_{1}$ is
mostly flat, $A_{2}$ is a monopole-type form factor and $V$ is a
dipole-type one \cite{YYK}. The values of the form factors
$A_1^{BK^*},A_2^{BK^*}$ and $V^{BK^*}$ at $q^2=0$ and
$q^2=m^2_{J/\psi}$ in various form-factor models are shown in
Table II.

\begin{table}[ht]
\caption{Form factors $A_1^{BK^*},A_2^{BK^*}$ and $V^{BK^*}$ at
$q^2=0$ and  $q^2=m^2_{J/\psi}$ in various form-factor models. }
\begin{center}
\begin{tabular}{l l l l l l l l l}
&  BSWI & BSWII & LF &  NS  & Yang & BB  & MS & YYK  \\
\hline
 $A^{BK^*}_1(0)$  & 0.33 & 0.33 & 0.26 & 0.30 & 0.18 & 0.34 & 0.36 & 0.49 \\
 $A^{BK^*}_1(m^2_{J/\psi})$  & 0.45 & 0.45 & 0.37 & 0.39 & 0.24 & 0.43 & 0.43 & 0.49 \\
 $A^{BK^*}_2(0)$  & 0.33 & 0.33 & 0.24 & 0.30 & 0.17 & 0.28 & 0.32 & 0.30 \\
 $A^{BK^*}_2(m^2_{J/\psi})$  & 0.46 & 0.63 & 0.43 & 0.48 & 0.31 & 0.45 & 0.50 & 0.42\\
 $V^{BK^*}(0)$ & 0.37 & 0.37 & 0.35 & 0.30 & 0.21 & 0.46 & 0.44 & 0.39 \\
 $V^{BK^*}(m^2_{J/\psi})$ & 0.55 & 0.82 & 0.42 & 0.51 & 0.40 & 0.86 & 0.77
& 0.87 \\
\end{tabular}
\end{center}
\end{table}

Among the eight form-factor models, only a few of them are
consistent with the lattice calculations at large $q^2$,
constraint from $B\to \phi K^*$ at low $q^2$ and the constraint
from heavy quark symmetry for the form-factor $q^2$ dependence.
The BSWI model assumes a monopole behavior (i.e. $n=1$) for all
the form factors. However, this is not consistent with heavy quark
symmetry for heavy-to-heavy transition. The BSWII model takes the
BSW model results for the form factors at zero momentum transfer
but makes a different ansatz for their $q^2$ dependence, namely a
dipole behavior (i.e. $n=2$) is assumed for the form factors
$F_1,~A_0,~A_2,~V$, motivated by heavy quark symmetry, and a
monopole dependence for $F_0,A_1$. However, the equality of the
form factors $A_1^{BK^*}$ and $A_2^{BK^*}$ at $q^2=0$ is ruled out
by recent measurements of $B\to\phi K^*$ decays \cite{CYvv}.
Lattice calculations of $V^{BK^*}$, $A_0^{BK^*}$ and $A_1^{BK^*}$
at large $q^2$ \cite{UKQCD} in conjunction with reasonable
extrapolation to $q^2=m^2_\J$ indicate that $V^{BK^*}(m_\J^2)$ is
of order 0.70-0.80.

The parameters $\tilde a^h(\J K^*)$ defined by
 \be
 \tilde a^h(\J K^*)=a^h_2+a^h_3+a^h_5+a^h_7+a^h_9
 \en
are calculated using Eq. (\ref{ai}) and their results are shown in
Table III. Since the penguin parameters $a_{3,5,7,9}^h$ are small,
in practice we have $\tilde a^h\approx a_2^h$. Note that $\tilde
a^0_2$ and $\tilde a_2^-$ are independent of the parameter
$\rho_H$ introduced in Eq. (\ref{logdiv}); that is, they are
infrared safe. Since $h_-$ is quite small due to the compensation
between the $A_1^{BK^*}$ and $V_1^{BK^*}$ terms and
$f^\pm_{II(3)}$ is inversely proportional to $h_-$, $\tilde a^-$
becomes more sensitive than $\tilde a^+$ to the form-factor model
chosen.

\begin{table}[htb]
\caption{The calculated parameters $\tilde a^h(\J K^*)$
$(h=0,+,-)$ for $B\to \J K^*$ decay in QCD factorization using
various form-factor models for $B-K^*$ transition. The
experimental results for $\tilde a^h(\J K^*)$ are obtained using
the averaged branching ratio of $B\to \J K^*$ measured by BaBar,
Belle and CLEO in conjunction with the central values of the BaBar
measurement for the spin amplitudes $|\hat A_{0,\bot,\|}|^2$. Only
the central values of $\tilde a^h_{\rm expt}$ are shown here.}
\begin{center}
\begin{tabular}{ l c l c l c l}
& $\tilde a^0$ & $|\tilde a^0|_{\rm expt}$ & $\tilde a^+$ &
$|\tilde a^+|_{\rm expt}$
& $\tilde a^-$ & $|\tilde a^-|_{\rm expt}$ \\
\hline
BSWI & $0.11-i0.06$ & 0.19 & $0.16-i0.05$ & 0.18 & $-0.01+i0.05$ & 0.06 \\
BSWII & $0.15-i0.06$ & 0.25 & $0.14-i0.05$ & 0.15 & $-0.07+i0.05$ & 0.14 \\
LF & $0.14-i0.06$ & 0.25 & $0.19-i0.05$ & 0.23 & $-0.02+i0.05$ & 0.07\\
NS & $0.14-i0.06$ & 0.25 & $0.18-i0.05$ & 0.20 & $-0.03+i0.05$ & 0.08  \\
Yang & $0.23-i0.06$ & 0.43 & $0.25-i0.05$ & 0.30 & $-0.16+i0.05$ & 0.20 \\
BB & $0.12-i0.06$ & 0.20 & $0.14-i0.05$ & 0.16 & $-0.15+i0.05$ & 0.23 \\
MS & $0.13-i0.06$ & 0.22 & $0.14-i0.05$ & 0.16 & $-0.07+i0.05$ & 0.14 \\
YYK & $0.09-i0.06$& 0.16 & $0.13-i0.05$ & 0.15 & $-0.06+i0.05$ & 0.12 \\
\end{tabular}
\end{center}
\end{table}

From the experimental measurement of spin amplitudes, it is ready
to extract the parameters $\tilde a^h$ in various form-factor
models. We use the averaged decay rate $\Gamma(B\to\J
K^*)=(5.34\pm 0.23)\times 10^{-16}$ GeV obtained from Eqs.
(\ref{BRcharge}) and (\ref{BRneutral}) and the central values of
the spin amplitudes measured by BaBar \cite{BaBar} as an
illustration:
 \be
&& |\hat A_0|^2=0.597\pm0.028\pm0.024, \quad |\hat
 A_\bot|^2=0.160\pm0.032\pm0.014, \non \\
 && |\hat A_\||^2=0.243\pm0.034\pm0.017.
 \en
Then $\tilde a^0$ can be determined from $\Gamma_L(B\to\J
K^*)=\Gamma(B\to\J K^*)\times|\hat A_0|^2$ and likewise for
$\tilde a^\pm$. The results are shown in Table III. It is evident
the ``experimental" values of $\tilde a^h$ are polarization
dependent: $|\tilde a^0|>|\tilde a^+|>|\tilde a^-|$, whereas the
present QCD factorization calculation yields $|\tilde a^+|>|\tilde
a^0|>|\tilde a^-|$.

Normalized spin amplitudes and their phases in $B\to J/\psi K^*$
decays calculated in various form-factor models using QCD
factorization are exhibited in Table IV, where the unknown
parameter $\rho_H$ in Eq.~(\ref{logdiv}) is taken to be real and
unity. For comparison, we also carry out the analysis in the
partial wave basis as the phases of $S$, $P$ and $D$ partial wave
amplitudes are the ones directly related to the long-range final
state interactions. We see from the Tables that the predicted
$|\hat A_0|^2$, $|D|^2$, and branching ratios are too small,
whereas $|\hat A_\bot|^2=|P|^2$ is too large. It is also clear
that a non-trivial phase $\phi_\|$ deviated from $-\pi$ is seen in
some form-factor models, but it is still too small compared to the
BaBar measurement. Nevertheless, a large phase $\phi_\|$ as
implied by BaBar can be achieved by adjusting the phase of the
complex parameter $\rho_H$, but admittedly it is rather arbitrary.
In other words, the present QCD factorization calculation cannot
say something definite for the phase $\phi_\|$. The partial wave
decompositions $S,P$ and $D$ corresponding to the relative orbital
angular momentum $L=0,1,2$ between $J/\psi$ and $K^*$ uniquely
determine the spin angular momentum. Our results are difficult to
account for the observation $|S|^2:|D|^2:|P|^2\simeq 3.5:1:1$ from
recent Babar and Belle measurements.

\begin{table}[htb]
\caption{Normalized spin amplitudes and their phases (in radians)
in $B\to J/\psi K^*$ decays calculated in various form-factor
models using QCD factorization. The branching ratios given in the
Table are for $B^+\to \J K^{*+}$. For comparison, experimental
results form CDF, CLEO, BaBar and Belle are also exhibited.
 }
\begin{center}
\begin{tabular}{l c c c c c c}
& $|\hat A_0|^2$ &  $|\hat A_\bot|^2$ &  $|\hat A_\||^2$ &
$\phi_\bot$ & $\phi_\|$ & ${\cal B}(10^{-3})$ \\ \hline
 BSWI & 0.43 & 0.33 & 0.24 & $-3.05$ & $-2.89$  & 0.76 \\
 BSWII & 0.38 & 0.36 & 0.26 & 3.13 & $-3.12$ & 0.73 \\
 LF & 0.41 & 0.34 & 0.25 & $-3.09$ & $-2.95$ & 0.69 \\
 NS & 0.40 & 0.34 & 0.25 & $-3.10$ & $-2.99$ & 0.70 \\
 Yang & 0.38 & 0.36 & 0.25 & $-3.12$ & $-3.11$ & 0.64 \\
 BB & 0.41 & 0.34 & 0.25 & $-3.04$ & $-3.05$ & 0.77 \\
 MS & 0.40 & 0.35 & 0.25 & $-3.08$ & $-3.05$ & 0.75 \\
YYK & 0.44 & 0.32 & 0.23 & $-2.99$ & $-2.95$ & 0.84  \\
 \hline
 CLEO \cite{CLEO}& $0.52\pm 0.08$ & $0.16\pm 0.09$ & $0.32\pm 0.12$& $-3.03\pm 0.46$ &
 $-3.00\pm 0.37$ & $1.41\pm 0.31$ \\
 CDF \cite{CDF}& $0.59\pm 0.06$ & $0.13^{+0.13}_{-0.11}$ & $0.28\pm0.12$
 & $-2.58\pm 0.54$ & $-2.20\pm0.47$  &   \\
 BaBar \cite{BaBar} & $0.60\pm 0.04$ & $0.16\pm 0.03$ & $0.24\pm 0.04$ &
 $-2.97\pm 0.17$ & $-2.50\pm 0.22$ & $1.37\pm0.14$ \\
 Belle \cite{Belle} & $0.60\pm0.05$ & $0.19\pm0.06$ & $0.21\pm 0.08$ & $-3.15\pm0.21$ &
 $-2.86\pm0.25$ & $1.29\pm0.14$ \\
\end{tabular}
\end{center}
\end{table}

\begin{table}[htb]
\caption{Normalized partial wave amplitudes and their phases (in
radians) in $B\to J/\psi K^*$ decays calculated in various
form-factor models using QCD factorization and fitted from the
data, where $\phi_P={\rm arg}(PS^*)$, $\phi_D={\rm arg}(DS^*)$ and
there exists a phase ambiguity: $\phi_D  \to -\phi_D$ and
 $\phi_P \to \pm\pi-\phi_P$.}
\begin{center}
\begin{tabular}{l c c c c c c}
& $|S|^2$ &  $|P|^2$ &  $|D|^2$ & $\phi_P$ & $\phi_D$ \\
\hline
 BSWI & 0.60 & 0.33 & 0.07 & $-0.04$ & $2.75$  \\
 BSWII & 0.60 & 0.36 & 0.04 & 0.02 & $3.10$  \\
 LF & 0.60 & 0.34 & 0.06 & $-0.05$ & $2.80$  \\
 NS & 0.60 & 0.34 & 0.06 & $-0.05$ & $2.86$ \\
 Yang & 0.59 & 0.36 & 0.05 & $0.002$ & $3.07$ \\
 BB & 0.60 & 0.34 & 0.06 & $0.05$ & $2.99$  \\
 MS & 0.60 & 0.35 & 0.05 & $0.01$ & $2.97$ \\
YYK & 0.60 & 0.32 & 0.08 & $0.05$ & $2.85$  \\
 \hline
 CLEO \cite{CLEO}& $0.77\pm 0.19$  & $0.16\pm 0.09$ & $0.07\pm 0.03$ & $0.04\pm 0.59$ &$2.9\pm 0.59$
 \\
 CDF \cite{CDF}& $0.61\pm 0.34$ & $0.13^{+0.13}_{-0.11}$ & $0.26\pm0.20$
 & $0.10\pm 0.34$ & $2.17\pm 0.34$    \\
 BaBar \cite{BaBar} & $0.65\pm 0.13$ & $0.16\pm 0.03$ & $0.19\pm 0.10$ &
 $-0.13\pm 0.21$ & $2.44\pm 0.21$ \\
 Belle \cite{Belle} & $0.66\pm 0.14$ & $0.19\pm0.06$ & $0.15\pm 0.03$& $-0.14\pm 0.29$
 &$2.80\pm 0.29$
   \\
\end{tabular}
\end{center}
\end{table}

There are several major theoretical uncertainties in the
calculation: $B-K^*$ form factors, the twist-3 LCDAs of $K^*$ at
the scale $\mu_h$ and the infrared divergences occurred in twist-2
and twist-3 contributions. It has been advocated that Sudakov form
factor suppression may alleviate the soft divergence \cite{Keum}.
Hence, we have studied Sudakov effects explicitly and the detailed
results will be presented in a future publication. When partons in
the meson carry the transverse momentum through the exchange of
gluons, the Sudakov suppression effect will be naturally generated
due to large double logarithms ${\rm exp}[-{\alpha_s C_F \over 4
\pi} \ln^2({Q^2\over k_{\bot}^2})]$, which will suppress the
long-distance contributions in the small $k_{\bot}$ region and
give a sizable average $\la k_{\bot}^2\ra \sim \bar{\Lambda} m_B$,
where $\bar\Lambda=m_B-m_b$. This can resolve the singularity
problem occurring at the end point.  Basically, there is no
Sudakov suppression in the vertex correction since the end-point
singularity in the hard kernel is cancelled in the convolution.
However, for the hard spectator interaction, we can have large
Sudakov suppression effects at the end point since there are
sizable $\la k_{\bot}^2 \ra$ contributions in the propagators.
Especially, the end-point singularities without $k_{\bot}$ do not
compensate in the twist-3 contributions. We find that $\tilde
a_2^{0}$ is suppressed whereas $\tilde a_2^{-}$ is enhanced by the
Sudakov effect and conclude that Sudakov suppression cannot help
to solve the discrepancy between theory and experiment.

\section{Discussions and Conclusions}
The hadronic decay $B\to\J K^*$ is analyzed within the framework
of QCD factorization. The spin amplitudes $A_0$, $A_\|$ and
$A_\bot$ in the transversity basis and their relative phases are
studied using various different form-factor models for $B-K^*$
transition. The effective parameters $a_2^h$ for helicity
$h=0,+,-$ states receive different nonfactorizable contributions
and hence they are helicity dependent, contrary to naive
factorization where $a_2^h$ are universal and polarization
independent. QCD factorization breaks down even at the twist-2
level for transverse hard spectator interactions. Although a
nontrivial strong phase for the $A_\|$ amplitude can be achieved
by adjusting the phase of an infrared divergent contribution, the
present QCD factorization calculation cannot say anything definite
about the phase $\phi_\|$. In QCD factorization we found that
$a_2^0$ and $a_2^-$ are infrared safe.

Unfortunately, our conclusion is somewhat negative. the
longitudinal parameter $a_2^0$ calculated by QCD factorization
which is of order 0.15 in magnitude is not large enough to account
for the observed decay rates and the fraction of longitudinal
polarization. In QCD factorization, the ratio $R$ of vector meson
to pseudoscalar production is close to unity  with large
uncertainties arising from the chirally enhanced and infrared
sensitive contributions to $B\to J/\psi K$~\cite{CYJpsiK}. (In the
naive factorization approach, $R$ ranges from 1.3 to 4.2
\cite{Cheng:1999kd}, but it is difficult to account for $R$,
$\Gamma_L/\Gamma$, and $|P|^2$ simultaneously.) This is mainly
ascribed to the smallness of $a_2^0$. It is instructive to compare
$a_2^0(\J K^*)$ in $B\to \J K^*$ decay with $a_2(\J K)$ in $B\to
\J K$. It is found in \cite{CYJpsiK} that $a_2(\J
K)=0.19^{+0.14}_{-0.12}$ for $|\rho_H|\leq 1$ and that twist-2 as
well as twist-3 hard spectator interactions are equally important.
As for $a_2^0(\J K^*)$, it is dominated by twist-2 hard spectator
interactions. We have studied Sudakov form factor suppression on
end-point singularities and found that it does not help to solve
the discrepancy between theory and experiment.

Since the predicted $a_2^0$ in QCD factorization is too small
compared to experiment, one may explore other effects that have
not been studied. One possibility is that soft final-state
interactions (FSIs) may enhance $a_2^0$
substantially~\cite{D0pi0}. A recent observation of $\ov B^0\to
D^{0(*)}\pi^0$ decay by Belle \cite{DpiBelle} and CLEO
\cite{DpiCLEO} indicates $a_2(D^{(*)}\pi)\sim 0.40-0.55$  much
larger than the naive value of order 0.25. It is thus conceivable
that some sort of inelastic FSIs could make substantial
nonperturbative contributions to $a_2^0$. The other possibility
arises from the gluon component in the $K^*$ wave function.
Consider the diagram in which one of outgoing charmed quarks emits
a hard gluon before they form the $J/\psi$ meson and the gluon
fragments into a parton of the $K^*$ meson. Neglecting the charmed
quark mass, because the charmed quark's helicity is conserved in
the strong interaction, this gluon has zero helicity, i.e., it is
longitudinally polarized. Following the same argument right after
Eq.~(\ref{Hii}), the hybrid $K^*$ will make a  contribution to
$H_0$ and $H_+$. Although this amplitude is suppressed by order of
$\Lambda_{\rm QCD}/m_b$ owing to the presence of an additional
propagator compared to the leading diagram, it is enhanced by the
large Wilson coefficient $c_1$ and hence cannot be ignored. A
similar mechanism can also give a contribution to the $B\to J/\psi
K$ mode but it is difficult to make a quantitative estimate since
the chirally enhanced twist-3 contribution is still quite
uncertain. Good candidates to search for evidence of this effect
are $B\to \rho^0 \rho^0, \rho^0 \omega, \omega \omega$. Without
taking into account the hard gluon emission, the branching ratios
of these decays which are color suppressed and dominated by $b\to
d$ penguin contributions are of order
$10^{-7}$\cite{CCTY,HY,CYvv}. Nevertheless, they can receive large
contributions, proportional to $c_1$ at the amplitude level, from
the hard gluon emission mechanism so that the branching ratios
become $10^{-6}\sim 10^{-5}$.

\vskip 1.0cm
 \noindent {\it Note added}: We learned the paper by X.S. Nguyen and X.Y.
Pham (NP) (hep-ph/0110284, v2) in which a similar analysis in QCD
factorization was carried out. However, their results differ from
ours in some aspects: (i) There are some discrepancies between
Eqs. (\ref{fI}-\ref{fh}) in the present paper and Eqs. (36) and
(37) of NP. Also the expression of $F_{II}^\pm$ given by Eq. (39)
of NP originally derived in \cite{CYvv} is valid only for two {\it
light} vector mesons in the final state. It will get some
modifications for heavy $\J$. It should be stressed  that Eq. (28)
adopted by NP for describing LCDAs works only for a light vector
meson, but not for a heavy meson like $\J$. (ii) For hard
spectator interactions, we have considered contributions from
leading wave functions of $B$ and $\J$ and twist-3 DAs of $K^*$
[see Eq. (\ref{fII3})], which are absent in NP. Also we have taken
into account the relevant scale $\mu_h=\sqrt{\Lambda_h m_b}$ for
hard spectator interactions. (iii) Unlike NP we did not consider
the higher twist expansion for the $\J$ wave function. The twist
expansion of LCDAs is applicable for light mesons but it is
problematic for heavy mesons like $\J$. Note that although twist-3
$\J$ contributions to hard spectator interactions were considered
by NP, they did not consistently compute the twist-3 effects of
$\J$ in vertex corrections.

\vskip 3.0cm \acknowledgments  H.Y.C. wishes to thank Physics
Department, Brookhaven National Laboratory and C.N. Yang Institute
for Theoretical Physics at SUNY Stony Brook for their hospitality.
Y.Y.K. would like to thank Physics Department, KIAS in Korea and
C.W. Kim for their hospitality. This work was supported in part by
the National Science Council of R.O.C. under Grant Nos.
NSC90-2112-M-001-047, NSC90-2112-M-033-004 and
NSC90-2811-M-001-019.


\end{document}